\documentclass[aps,prd,twocolumn,floatfix,nofootinbib]{revtex4-2}
\usepackage{amsmath,amssymb,mathtools,bm,mathrsfs}
\usepackage{graphicx}
\usepackage{dcolumn}
\usepackage{hyperref}
\usepackage{xcolor}
\hypersetup{colorlinks=true,citecolor=blue,linkcolor=blue,urlcolor=blue}
\graphicspath{{figures/}}

\newcommand{\rh}{r_{\rm h}}
\newcommand{\Tc}{T_{\rm c}}
\newcommand{\Pc}{P_{\rm c}}
\newcommand{\Gc}{G_{\rm c}}
\newcommand{\Lc}{\mathcal{L}}
\newcommand{\Fc}{\mathcal{F}}
\newcommand{\Dc}{\mathcal{D}}

\begin{document}
\title{Complex phase diagram and thermodynamic supercriticality of five-dimensional four-charge AdS black holes in $STU-W^2U$ gauged supergravity}

\author{Di Wu$^{1,2}$}
\email{Contact author: wdcwnu@163.com, d277wu@uwaterloo.ca}
%https://orcid.org/0000-0002-2509-6729

\author{Xiao-Dan Zhu$^1$}
\email{Contact author: zxdcwnu@163.com}

\affiliation{$^1$School of Physics and Astronomy, China West Normal University,
Nanchong, Sichuan 637002, People's Republic of China \\
$^2$Department of Physics and Astronomy, University of Waterloo, Waterloo, Ontario N2L 3G1, Canada}

\date{\today}

\begin{abstract}
In this paper, we investigate the complex phase structure and thermodynamic supercriticality of static four-charge anti-de Sitter (AdS) black holes in five-dimensional $STU-W^2U$ gauged supergravity, with all four physical electric charges held fixed. A Van der Waals-like first-order small/large black hole phase transition is found to terminate at a critical point. By analytically continuing the fixed-charge thermodynamic relations, we trace the two real spinodal singularities through criticality into a complex-conjugate pair in the supercritical region. The corresponding spinodal temperatures exhibit the characteristic $3/2$ near-critical scaling, connecting the subcritical temperature splitting with the imaginary parts of the supercritical complex temperatures. We construct the associated complex phase diagram and identify a Widom-type crossover line from the real-temperature projection of the complex spinodals. This line progressively deviates from the locus of real heat-capacity maxima as the pressure increases. At higher pressures, the real heat-capacity extrema eventually merge and disappear, while the complex spinodal singularities persist. Together, these results reveal a continuous connection between critical phase transitions and supercritical thermodynamics through complex singularity structures, providing further insight into the thermodynamic phase behavior of multicharge supergravity black holes.
\end{abstract}

\maketitle

\section{Introduction}
\label{sec:introduction}

Black hole thermodynamics has revealed a remarkably rich variety of phase
structures in gravitational systems \cite{NatRevPhys8-425}. In anti-de Sitter (AdS) spacetime,
these include the Hawking-Page transition and reentrant phase transitions
\cite{CMP87-577,PRD88-101502,JHEP0914080}, multiple coexistence points
\cite{CQG31-042001,JHEP1222117,JHEP0823028}, $\lambda$-line criticality
\cite{PRL118-021301}, and temperature-dependent thermodynamic topological
phase transitions \cite{JHEP0624213,PRD111-L061501,2607.07364}. An
especially intriguing feature of some black hole
phase diagrams is the termination of a first-order coexistence line at a
critical point. The Van der Waals-like small/large black hole transition
provides a prominent example of this behavior \cite{JHEP0712033,JHEP1112110,
CQG28-235017}, raising a natural question: what thermodynamic structures
characterize the black hole system in the supercritical region?

Above the critical point, AdS black holes can exhibit continuous
thermodynamic crossovers, reflected in pronounced variations of their
response functions. Extrema of thermodynamic response functions, such as
the heat capacity, have been used to identify characteristic supercritical
crossover curves, often associated with the Widom line
\cite{CPL42-101102,PRD112-064004}. Recent studies have further identified
two crossover boundaries that distinguish small-black-hole-like,
intermediate, and large-black-hole-like regimes in the supercritical
region \cite{Research9-1406,PLB876-140438}. These real-axis thermodynamic signatures
provide a useful characterization of supercritical crossovers and motivate
a closer examination of their connection to the singularity structure
associated with criticality.

Analytic continuation of black hole thermodynamic relations into the
complex plane provides a way to investigate the singularity structure
underlying supercritical crossovers \cite{PLB850-138528,EPJC86-1146,2608.09980}.
This approach is motivated by the
broader connection between complex-plane singularities and phase
transitions established in Lee-Yang theory \cite{PR87-404,PR87-410}.
A complex thermodynamic phase diagram has been constructed for charged
AdS black holes, revealing a Widom line that characterizes distinct
supercritical regimes \cite{PRL136-041402}. Subsequent studies have explored
complex phase structures in Born-Infeld and rotating AdS black holes
\cite{EPJC86-524,2610.24694}. Recent investigations of Kerr-Newman-AdS
black holes have further shown that complex spinodal singularities and
real thermodynamic response extrema characterize distinct aspects of
supercritical behavior \cite{2610.24978}.

The static four-charge AdS black hole constructed in the five-dimensional
$STU-W^2U$ gauged supergravity model \cite{JHEP0226252} provides an interesting
setting for examining the relationship between complex thermodynamic
singularities and real supercritical responses. Defined by the cubic
prepotential $\mathcal{V}=STU-W^2U=1$, this model extends the conventional
$STU$ model with an additional Abelian gauge field and nontrivial gauge-field
and scalar couplings. Furthermore, the fourth gauge field has been shown
to induce a splitting of black hole solutions into distinct branches in
a related construction \cite{EPJC86-635}. Previous thermodynamic investigations
of this four-charge AdS black hole have also revealed a new thermodynamic
topological subclass \cite{EPJC86-187}.

In this paper, we investigate the complex phase structure and thermodynamic
supercriticality of the static four-charge AdS black hole in the
five-dimensional $STU-W^2U$ model. Our main focus is on the connection
between the spinodal singularities associated with the first-order
small/large black hole transition and the real thermodynamic responses
beyond criticality. We identify the critical endpoint and trace the real
spinodal singularities into a complex-conjugate pair in the supercritical
region. Their real-temperature projections progressively deviate from the
locations of the heat-capacity maxima as the pressure increases.
Eventually, the real heat-capacity extrema merge and disappear, while the
complex spinodal singularities persist.

This paper is organized as follows. In Sec.~\ref{sec:background}, we introduce
the five-dimensional four-charge AdS black hole and its thermodynamic
properties, and analyze its critical point and real phase structure.
In Sec.~\ref{sec:complexroots}, we investigate the analytic continuation
of spinodal singularities into the complex plane and their near-critical
behavior. In Sec.~\ref{sec:phase}, we construct the complex phase diagram
and examine the supercritical thermodynamic responses. Finally,
Sec.~\ref{sec:conclusions} summarizes our main findings and discusses
possible directions for future research.

\section{Five-dimensional four-charge AdS black holes and real thermodynamics}
\label{sec:background}

\subsection{The $STU-W^2U$ model and its static four-charge AdS black hole solution}
\label{subsec:model}

We consider the static, nonextremal, four-charge asymptotically
anti-de Sitter black hole of five-dimensional $\mathcal N=2$ gauged
supergravity coupled to three vector multiplets.  The solution, including its AdS extension, was constructed in
Ref.~\cite{JHEP0226252}, and its thermodynamic properties were
studied in Ref.~\cite{EPJC86-187}. We briefly review the black hole
solution and its basic thermodynamic properties before investigating
its phase structure.

The scalar sector is specified by the cubic prepotential
\begin{align}
 \mathcal V&=\frac16 C_{IJK}X^I X^J X^K =X^1X^2X^3-(X^4)^2X^3\nonumber\\
 &\equiv STU-W^2U=1.
 \label{eq:prepot}
\end{align}
where $I,J,K=1,\ldots,4$, $C_{123}=1$ and
$C_{344}=C_{434}=C_{443}=-2$ (together with the symmetric
permutations).  With the conventions of Ref.~\cite{EPJC86-187}, the
bosonic action is
\begin{align}
I_{\mathrm{bos}}=&{1\over16\pi G_5}\int d^5x\sqrt{-g}\bigg\{R
 -3(\partial\varphi_1)^2-\alpha(\partial\varphi_2)^2 \nonumber \\
&-{(\partial\alpha)^2\over4\alpha(\alpha-1)}
 -{\alpha\over4}\left(e^{-2\varphi_1-2\varphi_2}F_1^2
+e^{-2\varphi_1+2\varphi_2}F_2^2\right) \nonumber \\
&-{e^{4\varphi_1}\over4}F_3^2
 -{2\alpha-1\over2}e^{-2\varphi_1}F_4^2
 -{\alpha-1\over2}e^{-2\varphi_1}F_1F_2 \nonumber \\
&+\sqrt{\alpha(\alpha-1)}\left(e^{-2\varphi_1-\varphi_2}F_1F_4
 +e^{-2\varphi_1+\varphi_2}F_2F_4\right)\nonumber\\
&+{1\over8}\epsilon^{\mu\nu\rho\sigma\lambda}
 \left(F^1_{\mu\nu}F^2_{\rho\sigma}A^3_\lambda
       -F^4_{\mu\nu}F^4_{\rho\sigma}A^3_\lambda\right) \nonumber \\
&-{2\over\ell^2}\left[\sqrt\alpha
 \left(e^{-\varphi_1-\varphi_2}+e^{-\varphi_1+\varphi_2}\right)
 +e^{2\varphi_1}\right]\bigg\}.
\label{eq:bosonic-action}
\end{align}
Here $R$ is the Ricci scalar, $G_5$ is the five-dimensional
Newton constant, $\ell$ is the AdS radius, and
$\epsilon^{\mu\nu\rho\sigma\lambda}$ is the five-dimensional
Levi-Civita tensor. The relation $F^I=dA^I$ defines the field strength
of the $I$th Abelian gauge field. The three independent scalar fields
$(\varphi_1,\varphi_2,\alpha)$ may be parametrized by
\begin{align}
 X^1&=\sqrt\alpha\,e^{\varphi_1+\varphi_2},&
 X^2&=\sqrt\alpha\,e^{\varphi_1-\varphi_2},\nonumber\\
 X^3&=e^{-2\varphi_1},&
 X^4&=\sqrt{\alpha-1}\,e^{\varphi_1}.
\label{eq:scalars}
\end{align}

Following Ref.~\cite{EPJC86-187}, we introduce the functions
\begin{align}
 Z_i(r)&=1+{q_i\over r^2},\quad i=1,2,3,&
 Z_4(r)&={q_4\over r^2},\nonumber\\
 H(r)&=\bigl(Z_1Z_2-Z_4^2\bigr)Z_3,&
 B(r)&=Z_1Z_2-Z_4^2.
\label{eq:harmonics}
\end{align}
The line element takes the form
\begin{equation}
 ds^2=H^{1/3}\left[-{f(r)\over H(r)}dt^2
  +{dr^2\over f(r)}+r^2d\Omega_3^2\right],
 \label{eq:metric}
\end{equation}
Here $d\Omega_3^2$ denotes the metric on the unit three-sphere,
and the metric function is given by
\begin{equation}
 f(r)=1-{2m\over r^2}
 +{p_4^2+(w-1)q_4^2\over r^4}
 +{r^2\over\ell^2}H(r).
 \label{eq:lapse}
\end{equation}
The electric gauge potentials are
\begin{align}
 A_1&={p_1 Z_2-p_4 Z_4\over r^2B(r)}\,dt,&
 A_2&={p_2 Z_1-p_4 Z_4\over r^2B(r)}\,dt,\nonumber\\
 A_3&={p_3\over r^2Z_3}\,dt,&
 A_4&={q_4(p_2Z_1-p_1Z_2)\over(q_1-q_2)r^2B(r)}\,dt.
 \label{eq:gaugefields}
\end{align}
For completeness, the scalar profiles are fixed by
\begin{equation}
 \varphi_1={1\over6}\ln{Z_3^2\over B(r)},\qquad
 \varphi_2={1\over2}\ln{Z_2\over Z_1},\qquad
 \alpha={Z_1Z_2\over B(r)}.
 \label{eq:scalarprofiles}
\end{equation}
The charge parameters obey
\begin{align}
 p_1^2&=q_1^2+2mq_1+w q_4^2,\nonumber\\
 p_2^2&=q_2^2+2mq_2+w q_4^2,\nonumber\\
 p_3^2&=q_3^2+2mq_3+p_4^2+(w-1)q_4^2,
 \nonumber\\
 p_4&=q_4{p_1-p_2\over q_1-q_2},
 \label{eq:charge-constraints}
\end{align}
together with the consistency relation
\begin{align}
 p_3^2={}&p_4^2-q_4^2
       +(q_3-q_2)(q_3-q_1)\nonumber\\
 &+\frac{(q_3-q_2)p_1^2-(q_3-q_1)p_2^2}{q_1-q_2}.
 \label{eq:supplement-charge}
\end{align}
Here $m$ is the mass parameter, $q_I$ are auxiliary charge
parameters, $p_I$ determine the physical electric charges, and $w$
is an arbitrary constant. We set $w=1$ to simplify the parametrization.
At fixed physical electric charges, this choice does not alter the
thermodynamic phase structure on the same physical branch, as shown
explicitly in
Sec.~\ref{subsec:thermo}.

\subsection{Thermodynamic quantities and a fixed-charge parametrization}
\label{subsec:thermo}

The event horizon is the largest positive root of
$f(\rh)=0$. With $G_5=1$, the entropy, Hawking temperature,
and conserved electric charges read~\cite{EPJC86-187}
\begin{align}
 S&={\pi^2\over2}\rh^3\sqrt{H(\rh)},
 &T&={f'(\rh)\over4\pi\sqrt{H(\rh)}},
 \label{eq:ST}\\
 &Q_I={\pi\over4}p_I,&& I=1,\ldots,4.
 \label{eq:physicalQ}
\end{align}
The ADM mass and pressure read~\cite{EPJC86-187}
\begin{equation}
 M={\pi\over4}(3m+q_1+q_2+q_3),\qquad
 P={3\over4\pi\ell^2}.
 \label{eq:MP}
\end{equation}
The horizon electrostatic potentials are~\cite{EPJC86-187}
\begin{align}
 \Phi_1&=\left.{p_1Z_2-p_4Z_4\over r^2B(r)}\right|_{\rh},
 &\Phi_2&=\left.{p_2Z_1-p_4Z_4\over r^2B(r)}\right|_{\rh},\nonumber\\
 \Phi_3&=\left.{p_3\over r^2Z_3}\right|_{\rh},
 &\Phi_4&=\left.{q_4(p_2Z_1-p_1Z_2)\over(q_1-q_2)r^2B(r)}\right|_{\rh},
 \label{eq:potentials}
\end{align}
while the thermodynamic volume reads~\cite{EPJC86-187}
\begin{align}
 V={\pi^2\over6}\bigl[&3\rh^4
 +2(q_1+q_2+q_3)\rh^2 +(q_1+q_2)q_3 \nonumber \\
 &+q_1q_2-q_4^2\bigr].
 \label{eq:volume}
\end{align}
These quantities satisfy the extended first law and Smarr relation~\cite{EPJC86-187},
\begin{align}
 dM&=T\,dS+\sum_{i=1}^3\Phi_i\,dQ_i
       -2\Phi_4\,dQ_4+V\,dP,\label{eq:firstlaw}\\
 M&={3\over2}TS+\sum_{i=1}^3\Phi_iQ_i
       -2\Phi_4Q_4-VP.
 \label{eq:smarr}
\end{align}
The coefficient $-2$ in the fourth-charge work term follows
from the gauge-field normalization of the $STU-W^2U$ theory.
Throughout the phase analysis, we keep all four conserved electric
charges fixed, $dQ_I=0$ ($I=1,\ldots,4$). Since
$Q_I=\pi p_I/4$, this is equivalent to fixing the four $p_I$.
When the pressure and horizon state are varied, the auxiliary
parameters $q_I$ and $m$ are adjusted to satisfy the charge
constraints and the horizon equation.

To simplify the parametrization of the coupled charge sector
at fixed physical charges, we introduce the symmetric matrix
\begin{equation}
 \mathsf P=\begin{pmatrix}p_1&p_4\\p_4&p_2\end{pmatrix},\quad
 \lambda_{\pm}={p_1+p_2\pm\sqrt{(p_1-p_2)^2+4p_4^2}\over2},
 \label{eq:spectral}
\end{equation}
and set $\lambda_3=p_3$. We also introduce an auxiliary
nonextremality parameter $\mu$ and the functions $R_a(\mu)$,
associated with these charge-matrix eigenvalues, through
\begin{equation}
 m=\sqrt{\mu^2+p_4^2},\qquad
 R_a(\mu)=\sqrt{\mu^2+\lambda_a^2},
 \quad a\in\{-,+,3\},
 \label{eq:mu-radii}
\end{equation}
The charge constraints in Eq.~\eqref{eq:charge-constraints} can then be
recast into a matrix relation, from which the original auxiliary
parameters are reconstructed as
\begin{equation}
 \begin{pmatrix}q_1&q_4\\q_4&q_2\end{pmatrix}
 =\sqrt{\mu^2\hat I+\mathsf P^2}-m\hat I,
 \quad q_3=R_3-m.
 \label{eq:charge-matrix}
\end{equation}
Here $\hat I$ denotes the $2\times2$ identity matrix.
For a real physical configuration, the square roots in
Eqs.~\eqref{eq:mu-radii} and \eqref{eq:charge-matrix} are
taken on the branch continuously connected to the positive
real solution. This parametrization preserves all four fixed
physical charges, including the off-diagonal $p_4$ coupling.

We write $x=r_h^2$, introduce $y=x-m(\mu)$, and define
$h=\ell^{-2}=4\pi P/3$. The horizon condition is then exactly equivalent to
\begin{align}
 \Fc(y,\mu,h)&=y^2-\mu^2+h\Dc(y,\mu)=0,\label{eq:Fconstraint} \\
 \Dc(y,\mu)&=\bigl[y+R_-(\mu)\bigr]\bigl[y+R_+(\mu)\bigr]\bigl[y+R_3(\mu)\bigr],
 \label{eq:Ddef}
\end{align}
where the physical horizon radius is obtained from
\begin{equation}
 \rh^2=x=y+\sqrt{\mu^2+p_4^2}.
 \label{eq:rphys}
\end{equation}
On the horizon, $r^6H(r)=\Dc$, and the thermodynamic
quantities take the compact forms
\begin{align}
 S&={\pi^2\over2}\sqrt{\Dc},&
 T&={\Fc_y\over2\pi\sqrt{\Dc}},\nonumber\\
 M&={\pi\over4}(R_-+R_++R_3),&
 V&={\pi^2\over6}\Dc_y,
 \label{eq:effective-thermo}
\end{align}
Here subscripts denote partial derivatives with the other effective
variables held fixed. On a physical isobar, however, $y$ and $\mu$
are constrained by $\Fc=0$, so thermodynamic variations must be
taken along this constraint. Equations
\eqref{eq:Fconstraint}-\eqref{eq:effective-thermo} are the
starting point of our fixed-charge phase analysis.

For any other admissible $w$, the redefinitions
$\mu^2=m^2-p_4^2-(w-1)q_4^2$ and $y=r^2-m$
leave the matrix identity and the effective horizon constraint
unchanged at fixed $p_I$. Consequently, the thermodynamic phase
structure remains unchanged under variations of $w$
at fixed physical charges on the same physical branch.

The on-shell Gibbs free energy and fixed-charge heat capacity are
\begin{equation}
 G=M-TS,\qquad
 C_{P,Q_I}=T\left({\partial S\over\partial T}\right)_{P,Q_I}.
 \label{eq:Gcp}
\end{equation}

\subsection{The critical point and real phase structure}
\label{subsec:realphase}

At fixed pressure and electric charges, the spinodal points are
identified by extrema of the Hawking temperature along a physical
horizon branch. The critical point occurs when two such points merge.
Along a constant-$h$ branch, differentiating $\Fc(y,\mu,h)=0$
gives the total derivative
\begin{equation}
 \Lc=\left.{d\over d\mu}\right|_{h,Q_I,\Fc=0}
 =\partial_\mu-{\Fc_\mu\over\Fc_y}\partial_y,
 \label{eq:Loperator}
\end{equation}
where $\Fc_y\ne0$ on the branch used for the local parametrization.
Accordingly, the local critical conditions become
\begin{equation}
 \Fc=0,\qquad {dT\over d\mu}=0,\qquad {d^2T\over d\mu^2}=0,
 \label{eq:critical-conds}
\end{equation}
with both derivatives evaluated at fixed pressure and charges along
$\Fc=0$.

To investigate the phase structure, we consider the following
nontrivial four-charge configuration:
\begin{equation}
 (p_1,p_2,p_3,p_4)=\left(1,2,150,\frac1{\sqrt2}\right),
 \qquad w=G_5=1,
 \label{eq:charges-choice}
\end{equation}
All four electric charges are nonzero, and their pronounced hierarchy
allows us to explore the thermodynamic behavior of the coupled
four-charge system. In particular, this configuration exhibits a
physical critical point and a multibranch phase structure, making it
suitable for the subsequent analysis of complex thermodynamic
singularities. The corresponding critical values are
\begin{align}
 y_c&\simeq50.216006828,&
 \mu_c&\simeq92.159131491,\nonumber\\
 h_c&\simeq0.001301696262,&
 \Pc&\simeq0.000310757092,\nonumber\\
 \Tc&\simeq0.015656476839,&
 \Gc&\simeq117.573312.
 \label{eq:critical-values}
\end{align}
For these parameters the physical critical squared radius is
$r_{hc}^2=y_c+\sqrt{\mu_c^2+p_4^2}\simeq142.377851$.
With the critical values determined, we introduce the reduced
thermodynamic variables
\begin{equation}
 p={P\over\Pc},\quad t={T\over\Tc},\quad
 g={G\over\Gc},\quad c_p=\Pc^{3/2}C_{P,Q_I}.
 \label{eq:reduced}
\end{equation}
We now trace the physical horizon branches at fixed
$Q_I$ as the pressure is varied relative to $\Pc$.

\begin{figure}[t]
 \includegraphics[width=\columnwidth]{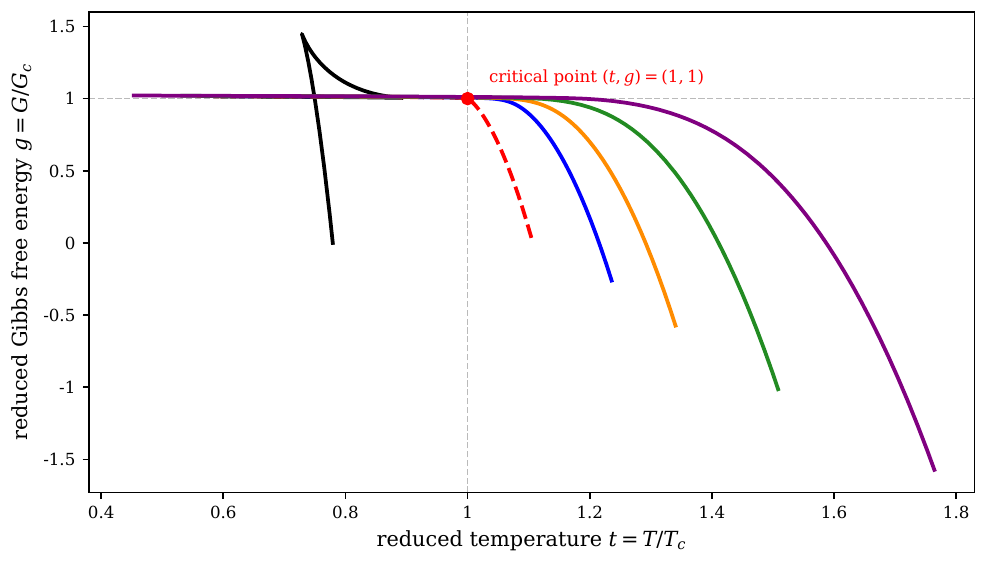}
 \caption{Reduced Gibbs free energy $g=G/G_c$ versus reduced
 temperature $t=T/T_c$ at fixed physical charges. The black
 solid, red dashed, blue solid, orange solid, dark-green solid,
 and purple solid curves correspond, respectively, to reduced
 pressures $p=P/P_c=0.5,1,1.2,1.35,1.6$, and $2$.
 The red point at $(t,g)=(1,1)$ marks the critical point;
 the gray dashed lines indicate $t=1$ and $g=1$.
 The subcritical multibranch structure contracts toward criticality,
 while the supercritical Gibbs curves are smooth.}
 \label{fig:gibbs}
\end{figure}

Figure~\ref{fig:gibbs} displays $g(t)$ for several reduced
pressures.  In the subcritical regime the free energy is
multivalued in temperature, with small- and large-black-hole
branches separated by an intermediate branch.  Equality of
the two competing Gibbs free energies determines the
first-order coexistence temperature.  The terminating cusps
of the multivalued curve are associated with the loss of local
stability of the respective metastable branches.  At the
critical pressure the two limiting points merge; for the
supercritical isobars shown the Gibbs curve no longer
exhibits the subcritical swallowtail structure.

While the Gibbs free energy determines the globally preferred
state among the competing black hole branches, the heat capacity
characterizes their local thermal stability. To display the heat-capacity
branches, we introduce the auxiliary dimensionless horizon coordinate
\begin{equation}
 u={y+\mu\over y_c+\mu_c}.
 \label{eq:u-aux}
\end{equation}
For $p_4\ne0$, the auxiliary coordinate $u$ differs from the
reduced physical squared horizon radius; their exact relation is
\begin{equation}
 {\rh^2\over r_{hc}^2}
 ={u(y_c+\mu_c)-\mu+\sqrt{\mu^2+p_4^2}
 \over y_c+\sqrt{\mu_c^2+p_4^2}}.
 \label{eq:u-to-rh}
\end{equation}

\begin{figure}[t]
 \includegraphics[width=\columnwidth]{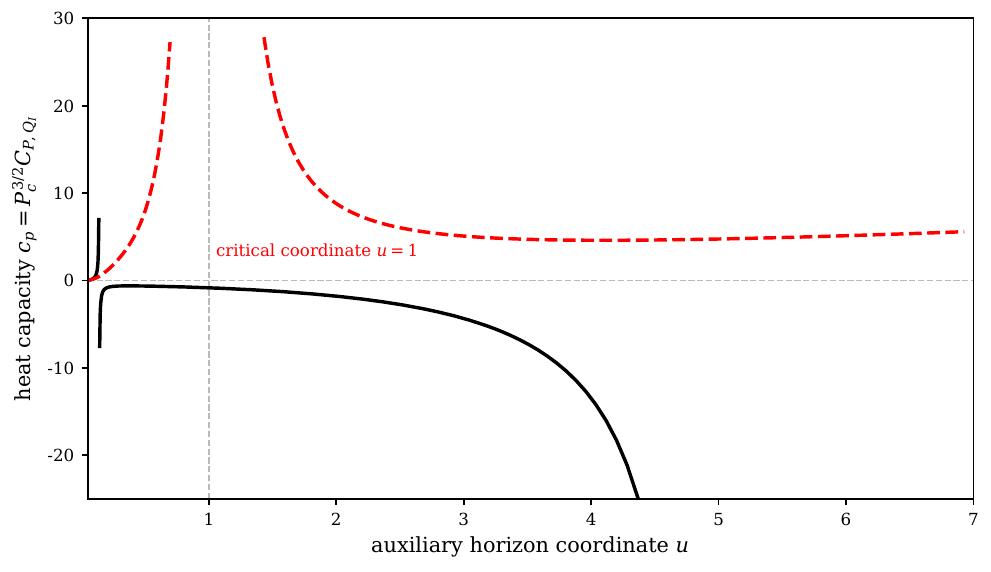}
 \caption{Reduced fixed-charge heat capacity
 $c_p=P_c^{3/2}C_{P,Q_I}$ versus the auxiliary horizon
 coordinate $u=(y+\mu)/(y_c+\mu_c)$. The black solid branches show $p=0.5$,
 and the red dashed branches show $p=1$. The vertical gray
 dashed guide indicates the critical coordinate $u=1$ and
 the horizontal guide $c_p=0$; the red annotation labels $u=1$.
 The two subcritical spinodal divergences delimit the unstable
 branch and merge at criticality.}
 \label{fig:cpbelow}
\end{figure}

As Fig.~\ref{fig:cpbelow} shows,
at $p=0.5$ the thermal response changes sign across two
spinodal divergences.  Positive heat capacity identifies
local stability with respect to thermal fluctuations in this
ensemble, whereas the intervening negative-heat-capacity
branch is locally unstable.  The spinodal temperatures are
therefore distinct from the coexistence temperature, which
is determined by the \emph{global} comparison of Gibbs
free energies.  The poles approach one another and coalesce
at the critical point, where the heat capacity becomes singular.

\begin{figure}[t]
 \includegraphics[width=\columnwidth]{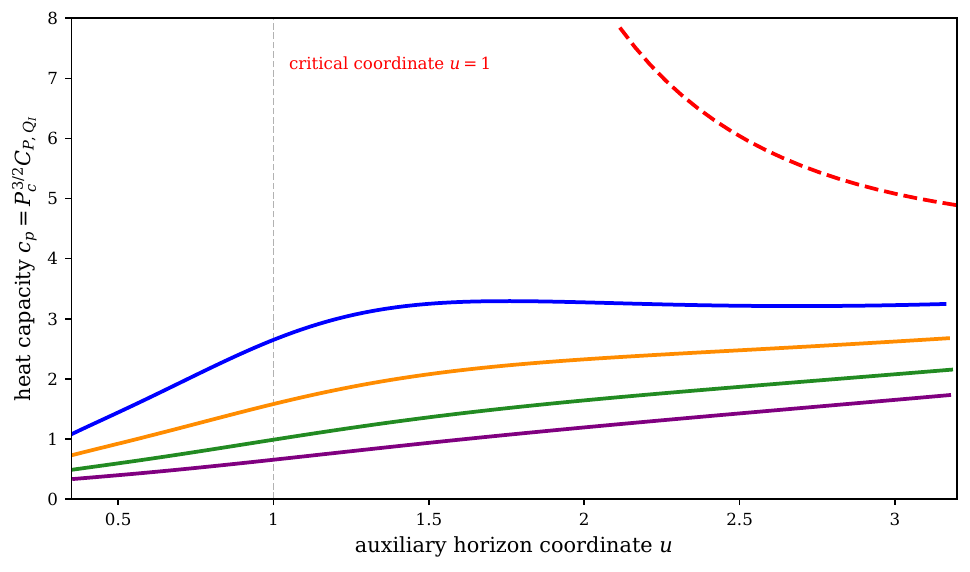}
 \caption{Reduced fixed-charge heat capacity in the critical
 and supercritical regime versus $u=(y+\mu)/(y_c+\mu_c)$. The red dashed,
 blue solid, orange solid, dark-green solid, and purple solid
 curves correspond, respectively, to $p=1,1.2,1.35,1.6$,
 and $2$. The vertical gray dashed line marks $u=1$.
 Away from criticality the real response is smooth; at $p=1.2$
 a local peak and shallow minimum remain in the plotted interval,
 whereas the higher-pressure curves shown are monotonic there.}
 \label{fig:cpabove}
\end{figure}

Above the critical pressure, the two real spinodal singularities
disappear, and the heat capacity becomes smooth along the physical
branch. Nevertheless, nontrivial local extrema can still survive.  In Fig.~\ref{fig:cpabove}, the $p=1.2$ curve
contains a peak followed by a shallow minimum, whereas
those extrema are absent from the displayed higher-pressure
curves.  Thus the disappearance of a real-axis divergence
need not coincide with the loss of all response-function
structure.  The natural question is how the two real
spinodal singularities evolve when the pressure is continued
through the critical value.  We address it by complexifying
the constrained thermodynamic equations.

\section{Complex thermodynamic singularities}
\label{sec:complexroots}

\subsection{Analytic continuation at fixed physical charges}
\label{subsec:continuation}

The complex extension of black-hole thermodynamic
singularities provides a useful description of the region
beyond the real critical point, as developed for charged
AdS and rotating AdS black holes in
Refs.~\cite{PRL136-041402,2610.24694}.  For the present
system the key technical step is to continue the horizon
condition and the response function \emph{at fixed four
physical charges}.  We allow $y$ and $\mu$ to be complex,
keep the four $p_I$ and the pressure $P$ real, and use
Eqs.~\eqref{eq:Fconstraint}-\eqref{eq:Ddef} with analytically
continued spectral radii $R_a(\mu)$.

At a regular point of the constraint manifold where
$\Fc_y\ne0$, Eq.~\eqref{eq:Loperator} gives the derivative
along the fixed-charge constrained branch. The heat capacity
can consequently be written as
\begin{equation}
 C_{P,Q_I}=T\,{\Lc(S)\over\Lc(T)}.
 \label{eq:cp-L}
\end{equation}
The complex spinodal equations are
\begin{equation}
 \boxed{\Fc(y,\mu,h)=0,\qquad\Lc(T)=0.}
 \label{eq:complex-spinodal}
\end{equation}
Their real positive-radius solutions reproduce the
subcritical spinodals of Sec.~\ref{subsec:realphase};
complex solutions yield the continuation of these
singularities away from the real thermodynamic branch.
We focus on the pair of complex singularity roots that are
continuously connected to the physical critical point.
The numerator $\Lc(S)$ is nonzero on the tracked complex
branch in the numerical scan, so the corresponding zeros of
$\Lc(T)$ remain poles of the analytically continued heat
capacity.

To display the complex singularity roots, we introduce the
dimensionless complex coordinate
\begin{equation}
 \zeta=\sqrt{\frac{y+\mu}{y_c+\mu_c}},
 \label{eq:zeta}
\end{equation}
This is the square-root continuation of the auxiliary coordinate
$u$ introduced in Sec.~\ref{subsec:realphase}, with
$\zeta^2=u$ and the critical point at $\zeta=1$.
The two signs $\pm\zeta$ represent the same underlying spinodal
solution and are displayed in Fig.~\ref{fig:complexroots} to show
the complete square-root coordinate structure.

\subsection{Root distribution and continuation across criticality}
\label{subsec:rootdistribution}

\begin{figure}[t]
 \includegraphics[width=\columnwidth]{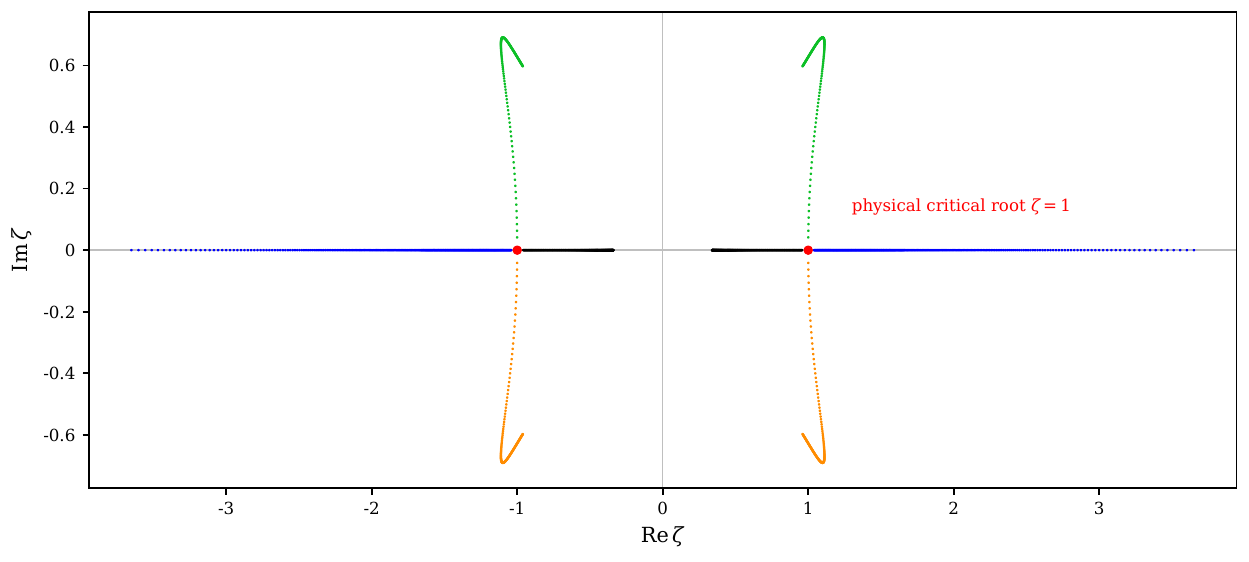}
 \caption{Complex singularity roots in the auxiliary $\zeta$
 plane obtained from Eq.~\eqref{eq:complex-spinodal}. Black
 and blue points represent the small- and large-black-hole
 spinodal root families for $0.2\le p\le0.999$, respectively.
 Green and orange points represent the upper and lower complex
 root families for $1.001\le p\le10$, respectively. Both
 algebraic signs $\pm\zeta$ are displayed. Red points mark
 $\zeta=\pm1$ at the critical pressure; the red annotation
 identifies the physical positive critical root $\zeta=1$.}
 \label{fig:complexroots}
\end{figure}

Figure~\ref{fig:complexroots} displays the root distribution
of Eq.~\eqref{eq:complex-spinodal}.  At $p<1$, the two
relevant solutions are real and correspond to the spinodal
limits of the small- and large-black-hole branches.  As
$p\to1^-$ they approach the common physical critical
solution.  When $p$ increases above unity, the continued
solutions move away from the real axis as a complex-conjugate
pair.  In this sense the critical point connects the real
spinodal structure to the complex singularities characterizing
the supercritical continuation.

In tracing the complex spinodal singularities, we continuously
follow the analytic branches of
$R_a(\mu)=\sqrt{\mu^2+\lambda_a^2}$ and the square root
$\sqrt{\Dc}$ appearing in the thermodynamic temperature,
starting from their physical real branches. This ensures a
consistent analytic continuation of both the singularity roots
and their associated complex temperatures.

The roots occupy a nontrivial trajectory in the complex
$\zeta$ plane.  Along the branch with positive real and
imaginary parts, the real component reaches a maximum near
$p\simeq1.88$, while $|\zeta|$ and the imaginary component
reach maxima near $p\simeq2.25$ and $p\simeq2.74$,
respectively.  The subsequent return of the trajectory is a
feature of the complex root geometry for the present fixed
charge choice.

\subsection{Near-critical relation between real and complex spinodals}
\label{subsec:scaling}

The connection between the real and complex spinodal roots can
be quantified near the critical pressure. We introduce
$\epsilon=p-1$ and $\xi=(\mu-\mu_c)/\mu_c$, where $\mu_c$ is
the critical value of the auxiliary nonextremality parameter.
Along the fixed-charge horizon constraint, the reduced temperature
can be regarded locally as a function $t(\xi,\epsilon)$.
Since its first and second derivatives with respect to $\xi$
vanish at criticality, the expansion takes the form
\begin{equation}
 t=1+A\epsilon+B\epsilon\xi+C\xi^3
       +O(\epsilon^2,\epsilon\xi^2,\xi^4),
 \label{eq:cusp}
\end{equation}
where the real coefficients $A$, $B$, and $C$ depend on the fixed
electric charges. Along an isobar, a spinodal root satisfies
\begin{equation}
 {\partial t\over\partial\xi}=0,
 \qquad
 B\epsilon+3C\xi^2+\cdots=0.
 \label{eq:local-spinodal}
\end{equation}
For the nondegenerate critical structure considered here,
$B,C\ne0$ and $B/C>0$, giving
\begin{equation}
 \xi_\pm=\pm\sqrt{-{B\epsilon\over3C}}+O(\epsilon).
 \label{eq:spinodal-xi}
\end{equation}
The two spinodal roots are real for $\epsilon<0$ and continue
as a complex-conjugate pair for $\epsilon>0$. Substitution into
Eq.~\eqref{eq:cusp} gives the subcritical spinodal temperatures
\begin{equation}
 t_\pm=1+A\epsilon\pm K(-\epsilon)^{3/2}+\cdots,
 \qquad \epsilon<0,
 \label{eq:subcrit-split}
\end{equation}
Here $t_+$ and $t_-$ denote the higher and lower spinodal
temperatures, respectively; their signs need not match those of
the auxiliary root labels $\xi_\pm$ in Eq.~\eqref{eq:spinodal-xi}.
The corresponding supercritical complex temperatures are
\begin{equation}
 t_*^\pm=1+A\epsilon\pm iK\epsilon^{3/2}+\cdots,
 \qquad \epsilon>0,
 \label{eq:supercrit-split}
\end{equation}
where
\begin{equation}
 K={2|B|\over3}\sqrt{B\over3C}>0.
 \label{eq:scaling-amplitude}
\end{equation}
Thus the $3/2$ exponent follows from the local structure
of a nondegenerate critical point. Moreover, the splitting
between the two real spinodal temperatures has amplitude $2K$,
whereas the magnitude of each complex-temperature imaginary
part has amplitude $K$. The values of these amplitudes depend
on the electric charges.

For the four-charge configuration in Eq.~\eqref{eq:charges-choice},
Eq.~\eqref{eq:scaling-amplitude} yields $K\simeq0.24594$.
The numerical singularity trajectories give
\begin{align}
 t_{\rm sp}^+-t_{\rm sp}^-&\simeq
        0.492\,(1-p)^{3/2},\qquad p<1,\nonumber\\
 |\operatorname{Im}t_*^\pm|&\simeq
        0.246\,(p-1)^{3/2},\qquad p>1.
 \label{eq:threehalf}
\end{align}
The numerical coefficients satisfy the factor-of-two relation
to the shown precision, in agreement with the local analytic
continuation. Their values characterize this fixed-charge
configuration. Related critical cusp scaling has been studied for
rotating charged AdS black holes \cite{2610.24978}; the present
calculation realizes this structure in the coupled four-charge
sector of five-dimensional gauged supergravity.

These results establish the connection between the real spinodal
singularities and their complex continuation across the critical
point. We next examine how these singularities, together with
the first-order coexistence line, determine the complex
thermodynamic phase structure and characterize the
supercritical regime.

\section{Complex phase diagram and thermodynamic supercriticality}
\label{sec:phase}

Having examined the complex thermodynamic singularities and their
near-critical behavior in Sec.~\ref{sec:complexroots}, we now investigate
how they are connected to the real phase structure and what they reveal
about thermodynamic supercriticality. We combine the first-order
coexistence curve, the real spinodal curves, and the continued complex
singularities to construct the phase diagram, and then compare the
resulting complex-singularity projection with real heat-capacity
features.

\begin{figure*}[t]
 \includegraphics[width=.91\textwidth]{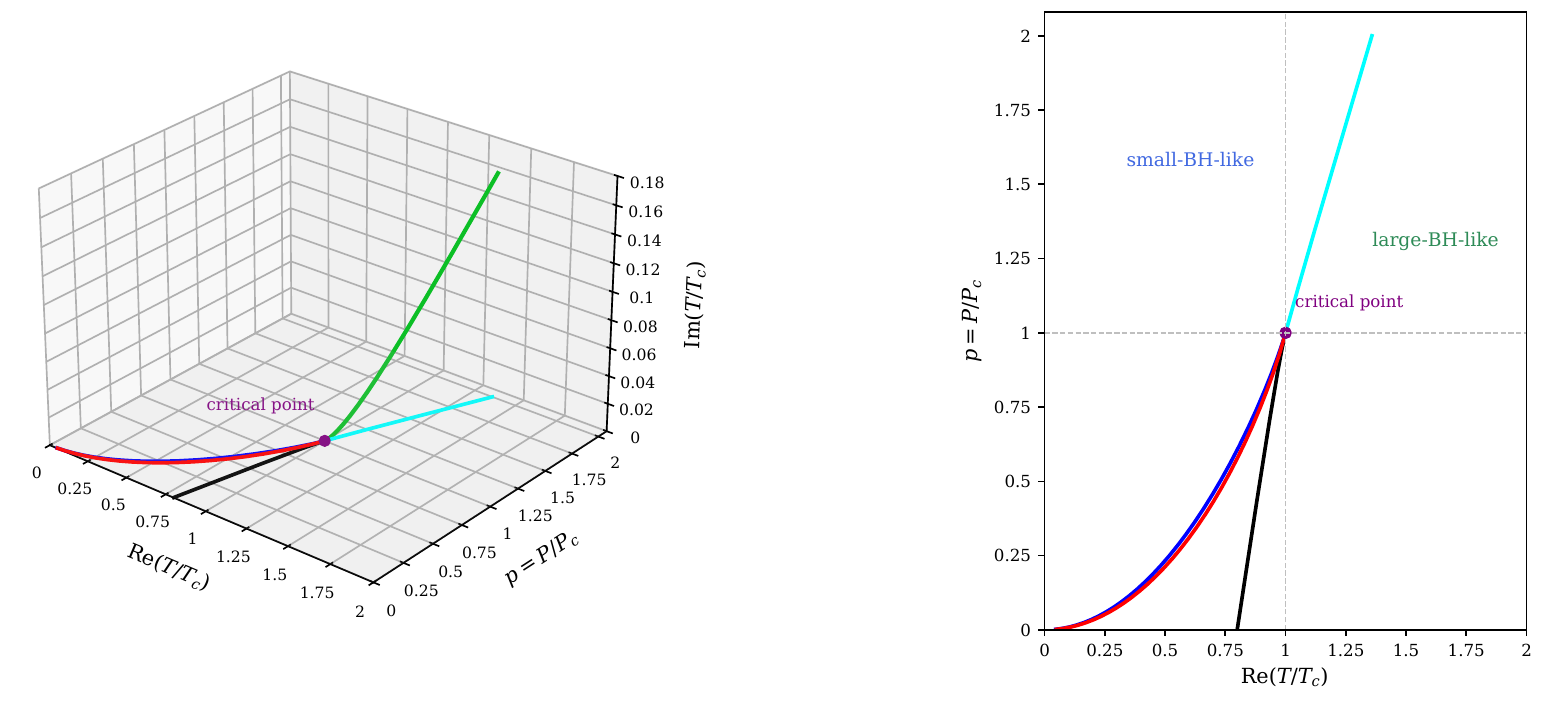}
 \caption{Complex thermodynamic phase diagram for fixed
 $(p_1,p_2,p_3,p_4)=(1,2,150,1/\sqrt2)$.
 Left: three-dimensional plot in
 $(\operatorname{Re}t,\,p,\,\operatorname{Im}t)$;
 right: projection onto the real temperature-pressure plane.
 In both panels the black and blue curves denote the small- and
 large-black-hole spinodals, the red curve denotes the first-order
 coexistence line, and the cyan curve denotes the Widom-type
 projection $t_W(p)=\operatorname{Re}t_*^+(p)$. The green
 curve in the left panel is the upper complex-singularity branch
 ($1.001\le p\le2$). The purple point is the critical point
 $(t,p)=(1,1)$, where the real curves meet the complex line.
 The right-panel labels identify the small-BH-like and
 large-BH-like supercritical regions. The lower complex-conjugate
 branch is omitted from the left panel for clarity.}
 \label{fig:phase}
\end{figure*}

\subsection{Coexistence, spinodal curves, and the complex phase line}
\label{subsec:diagram}

We first determine the real coexistence curve by requiring equal
temperatures and Gibbs free energies for the small black hole (SBH)
and large black hole (LBH) branches at the same pressure:
\begin{equation}
 T_{\rm SBH}(P)=T_{\rm LBH}(P),\qquad
 G_{\rm SBH}(P)=G_{\rm LBH}(P).
 \label{eq:coex}
\end{equation}
The coexistence curve satisfies the Clapeyron relation
\begin{equation}
 {dP\over dT}\bigg|_{\rm coex}
 =\frac{\Delta S}{\Delta V},
 \label{eq:clapeyron}
\end{equation}
where $\Delta S$ and $\Delta V$ are the differences between the
large- and small-black-hole states.

The two real spinodal curves follow from
Eq.~\eqref{eq:complex-spinodal} on the physical real horizon
branches. These curves meet at the critical endpoint of the
first-order coexistence line.

To describe the complex continuation above criticality, let
$T_*^\pm(p)$ denote the complex-conjugate singularity temperatures
obtained from Eq.~\eqref{eq:complex-spinodal} at the reduced pressure
$p=P/P_c$. Their reduced counterparts are
\begin{equation}
 t_*^\pm(p)=\frac{T_*^\pm(p)}{T_c},
 \label{eq:reduced-complex-temperature}
\end{equation}
where the plus sign labels the upper complex-temperature branch.
The continued singularities define two conjugate trajectories,
\begin{equation}
 \gamma_\pm(p)=\left(
    \operatorname{Re}t_*^\pm(p),\
    p,\
    \operatorname{Im}t_*^\pm(p)
 \right).
 \label{eq:complexline}
\end{equation}
Their common projection onto the real temperature-pressure plane
defines a Widom-type crossover line in the complex-singularity
prescription of Refs.~\cite{PRL136-041402,2610.24694}:
\begin{equation}
 t_W(p)=\operatorname{Re}t_*^+(p),\qquad p>1.
 \label{eq:widom}
\end{equation}
This projection is distinct in its definition from the locus of
extrema of a real response function, which we examine in
Sec.~\ref{subsec:widomheat}.

Figure~\ref{fig:phase} brings these structures together. The
coexistence line and the two real spinodal curves join at
$(t,p)=(1,1)$, while the complex singularity trajectories emerge
from the same point into the supercritical region. The right panel
shows their real-temperature projection alongside the subcritical
curves, making their common critical origin explicit.
Their near-critical connection follows the $3/2$ scaling
derived in Sec.~\ref{subsec:scaling}. The labels
``small-BH-like'' and ``large-BH-like'' in the supercritical
region describe crossover behavior, not distinct phases
separated by a first-order transition. Along a supercritical
isobar, increasing temperature corresponds to increasing black-hole
size, and $t_W(p)$ marks the crossover between these two regimes.

\subsection{Supercritical response and characteristic crossover temperatures}
\label{subsec:widomheat}

Having constructed the complex phase diagram, we next ask how its
projected characteristic line compares with the behavior of a real
thermodynamic response. In particular, we compare $t_W(p)$ with
the temperature $t_{C_P,\mathrm{max}}$ at which the fixed-charge
heat capacity has a local maximum. The former is determined by a
continued complex singularity, whereas the latter is defined by
an extremum of a finite real response. For the charge configuration
in Eq.~\eqref{eq:charges-choice}, their values are listed in
Table~\ref{tab:widom}.
\begin{table}[t]
\caption{The complex-singularity projection $t_W$ and the
 temperature $t_{C_P,\mathrm{max}}$ of the local real heat-capacity
 maximum at selected supercritical pressures, for the fixed charge
 configuration in Eq.~\eqref{eq:charges-choice}.}
\label{tab:widom}
\begin{ruledtabular}
\begin{tabular}{cccc}
$p$ & $t_W$ & $t_{C_P,\mathrm{max}}$ & $t_{C_P,\mathrm{max}}-t_W$\\
\hline
1.05 & 1.016956 & 1.017797 & 0.000841\\
1.1 & 1.034132 & 1.037716 & 0.003584\\
1.2 & 1.069068 & 1.088006 & 0.018938\\
\end{tabular}
\end{ruledtabular}
\end{table}
Near criticality both characteristic temperatures tend to unity,
but their difference grows with increasing pressure. Thus the
real-temperature projection of a complex singularity does not, in
general, coincide with a real-response peak, in line with related
discussions of distinct crossover prescriptions
\cite{2610.24978}.

More significantly, the two prescriptions differ in how far they
can be followed into the supercritical regime. For the selected
charge configuration, the local maximum and subsequent minimum of
the real heat capacity approach one another and merge near
\begin{equation}
 p_{\rm ext}\simeq1.228,\qquad u_{\rm ext}\simeq2.18,
 \label{eq:cpext}
\end{equation}
where
\begin{equation}
 \left({\partial C_{P,Q_I}\over\partial u}\right)_p=0,
 \qquad
 \left({\partial^2 C_{P,Q_I}\over\partial u^2}\right)_p=0.
 \label{eq:cpextconditions}
\end{equation}
Beyond this pressure, these local extrema are absent, consistent
with the monotonic curves at $p=1.35,1.6$, and $2$ in
Fig.~\ref{fig:cpabove}. In contrast, the complex singularity
trajectories and their projected line $t_W(p)$ can still be
followed beyond this point. This distinction shows that the
complex-singularity-based characteristic line is not contingent
on the persistence of the real heat-capacity peak.

\subsection{Extended continuation of complex roots and temperatures}
\label{subsec:farphase}

The persistence of the complex singularity line motivates a
closer examination of its behavior farther above criticality.
The auxiliary complex roots in Fig.~\ref{fig:complexroots} and
their images in the complex temperature plane need not have the
same dependence on pressure. Indeed, in the extended continuation
$1<p\le10$, the tracked $\zeta$ trajectory is nonmonotonic,
whereas both $\operatorname{Re}t_*^+$ and
$|\operatorname{Im}t_*^+|$ increase with $p$. For example,
the upper complex-temperature branch reaches
\begin{equation}
 t_*^+(p=10)\simeq3.874+1.516i,
 \label{eq:tfar}
\end{equation}
even though the auxiliary root trajectory has already turned
back toward smaller $\operatorname{Re}\zeta$ and
$\operatorname{Im}\zeta$. Thus the geometry of the auxiliary
complex-root trajectory does not by itself determine the behavior
of the corresponding complex thermodynamic temperature. To keep
the critical region legible, Fig.~\ref{fig:phase} displays only
the range $p\le2$.

These results establish a unified picture of the real and complex
phase structure of the five-dimensional four-charge AdS black hole.
In particular, the projected complex-singularity line differs from
the real heat-capacity peak line and persists even after the latter
disappears. This demonstrates that complex thermodynamic
singularities can characterize supercritical behavior beyond the
existence of local extrema in real thermodynamic response
functions.

\section{Conclusions and outlook}
\label{sec:conclusions}

In this work, we have investigated the real and complex thermodynamic
phase structure of the static four-charge AdS black hole in
five-dimensional $STU-W^2U$ gauged supergravity, with all four physical
electric charges held fixed. We identify a Van der Waals-like first-order
phase transition between small and large black holes, which terminates at
a critical point. The associated real spinodal singularities merge at
criticality and continue into the supercritical region as a
complex-conjugate pair. Near the critical point, the splitting of the
subcritical spinodal temperatures and the imaginary parts of the
supercritical complex temperatures exhibit the characteristic $3/2$
scaling, with their amplitudes differing by a factor of two.

By combining the coexistence curve, real spinodals, and their complex
continuation, we have constructed the complex thermodynamic phase diagram
of this four-charge black hole. The real-temperature projection of the
complex spinodal singularities defines a Widom-type crossover line, which
generally differs from the locus of real heat-capacity maxima. As the
pressure increases, the local maximum and minimum of the real heat
capacity eventually merge and disappear, while the complex spinodal
singularities continue to exist well beyond this point. These results
show that the singular structures associated with the critical phase
transition persist in the complex plane even after the corresponding
real thermodynamic response becomes smooth and monotonic.

The dependence of the critical behavior, complex spinodal trajectories,
and supercritical crossover characteristics on the four independent
electric charges deserves further investigation. It would also be
interesting to explore whether the solution-branch splitting found in
related four-charge black hole constructions is accompanied by distinct
critical and supercritical thermodynamic structures \cite{EPJC86-635}.
These investigations may further reveal how different charge configurations
and black hole solution branches influence the complex continuation of
critical singularities, thereby deepening our understanding of the
connection between critical phase transitions and supercritical
thermodynamics in multicharge supergravity black holes.

\begin{acknowledgments}
This work is supported by the National Natural Science Foundation of
China (NSFC) under Grants No. 12675077, No. 12205243 and No. 12375053, by the China
Scholarship Council (CSC), and by the Sichuan Science and Technology Program
under Grant No. 2026NSFSC0021.
\end{acknowledgments}

\end{document}